\documentclass[12pt]{article}
\usepackage{amsmath}
\usepackage{amsthm}
\usepackage{graphicx}
\usepackage{enumerate}
\usepackage{natbib}
\newtheorem{theorem}{Theorem}
\newtheorem{definition}{Definition}
\newtheorem{corollary}{Corollary}
\newtheorem{assumptions}{Assumptions}
\usepackage{bm}

\usepackage{url} 
\DeclareMathOperator*{\argmax}{arg\,max}

\usepackage[utf8]{inputenc} 
\usepackage[T1]{fontenc}    
\usepackage{hyperref}       
\usepackage{url}            
\usepackage{booktabs}       
\usepackage{amsfonts}       
\usepackage{nicefrac}       
\usepackage{microtype}      
\usepackage{xcolor}         
\usepackage{amsmath}
\usepackage[ruled,vlined]{algorithm2e}  

\usepackage{soul}
\usepackage{todonotes}

\newcommand{\blind}{1}

\begin{document}

\def\spacingset#1{\renewcommand{\baselinestretch}%
{#1}\small\normalsize} \spacingset{1}


\if1\blind
{
  \title{\bf Some models are useful, but when?: A decision-theoretic approach to choosing when to refit large-scale prediction models}
  \author{Kentaro Hoffman \hspace{.2cm}\\
    Department of Statistics, University of Washington\\
    Stephen Salerno \\
    Fred Hutch Cancer Center, Seattle\\
    Jeffrey Leek\thanks{
    This work was supported in part by NIH/NIGMS R35 GM144128 and the Fred Hutchinson Cancer Center J. Orin Edson Foundation Endowed Chair (SS and JTL), NIH/NIMH DP2 MH12}\\
    Fred Hutch Cancer Center, Seattle\\
    Tyler McCormick\thanks{
    This work was supported in part by NIH/NIMH DP2 MH122405, R01 HD107015, and P2C HD042828 (KH and TMC)} \\
    Department of Statistics and Department of Sociology, University of Washington}
  \maketitle
} \fi

\if0\blind
{
  \bigskip
  \bigskip
  \bigskip
  \begin{center}
    {\LARGE\bf Some models are useful, but when?: A decision-theoretic approach to choosing when to refit large-scale prediction models}
\end{center}
  \medskip
} \fi

\bigskip
\begin{abstract}
Large-scale prediction models using tools from artificial intelligence (AI) or machine learning (ML) are increasingly common across a variety of industries and scientific domains. Despite their effectiveness, training AI and ML tools at scale can cost tens or hundreds of thousands of dollars (or more); and even after a model is trained, substantial resources must be invested to keep models up-to-date. This paper presents a decision-theoretic framework for deciding when to refit an AI/ML model when the goal is to perform unbiased statistical inference using partially AI/ML-generated data. Drawing on portfolio optimization theory, we treat the decision of {\it recalibrating} a model or statistical inference versus {\it refitting} the model as a choice between ``investing'' in one of two ``assets.'' One asset, recalibrating the model based on another model, is quick and relatively inexpensive but bears uncertainty from sampling and may not be robust to model drift. The other asset, {\it refitting} the model, is costly but removes the drift concern (though not statistical uncertainty from sampling). We present a framework for balancing these two potential investments while preserving statistical validity.  We evaluate the framework using simulation and data on electricity usage and predicting flu trends.    


~\\~\\~\\
~\\~\\~\\

\end{abstract}
\noindent%
\vfill

\newpage
\spacingset{1.9} 

\section{Introduction}
\label{sec:intro}


With the rapid advancement of artificial intelligence (AI) and machine learning (ML), large-scale models are becoming standard tools in scientific research and industry settings to generate predictions of hard-to-measure outcomes. For example, AI/ML techniques applied to electronic health records can identify metastatic cancers~\citep{yang2022identification}, determine causes of death \citep{fan2024narratives,Byass2019,McCormick2016}, or provide targeted real-time early warning systems for adverse health events such as sepsis in emergent settings~\citep{wong2021external}. Transformer models such as AlphaFold can predict protein structures from easily obtained amino acid sequences \citep{jumper2021highly}. Deep learning of spatio-temporal dependencies in energy consumption data can forecast future consumption patterns ~\citep{somu2021deep}.  

Typically, these predicted outcomes are not the final goal, but intermediate sources of auxiliary information to use when estimating a particular target of inference (e.g., a population statistic, regression coefficient, or treatment effect). For example, global public health researchers estimate how cause-of-death varies by demographics using cause-of-death predicted using NLP techniques narratives from caretakers~\citep{fan2024narratives}. In pragmatic clinical trials, researchers predict outcomes that may have occurred between regularly scheduled clinic visits to estimate the effectiveness of interventions in everyday clinical practice ~\citep{gamerman2019pragmatic, williams2015pragmatic}.

This estimation step faces statistical, sampling-based uncertainty. Since the outcomes are predicted, valid inference also depends on how well the prediction model performs.  If the population for inference is the same as (or has a known relationship to) the population used to train the prediction model, then we can estimate, and correct for, the additional error introduced by the prediction model.  This emerging area \citep{wang2020methods, miao2023assumption, angelopoulos2023prediction, miao2024taskagnosticmachinelearningassistedinference, egami2024usingimperfectsurrogatesdownstream} is referred to in \citet{hoffman2024really} as ``Inference on Predicted Data.'' In many applied settings, however, the relationship between the population of interest for inference and the one used for training is opaque, and the practitioner must decide whether or not to use the model's predictions without full information about this relationship.
This paper addresses this increasingly ubiquitous conundrum. We present a decision-theoretic framework to decide when to refit an upstream prediction model when the goal is to perform unbiased downstream statistical inference on AI/ML-predicted outcomes. Drawing on portfolio optimization theory, we treat this problem as a choice between ``investing'' in one of two ``assets.''

Performing inference using predicted outcomes requires managing traditional, sampling-based, statistical uncertainty as well as uncertainty about how appropriate the prediction model is to the current data setting. Mismatch between the mapping of features to outcomes between one population and another is known as ``concept drift.'' Concept drift can be gradual or sudden, making it difficult to quantify without extensive data from the new population. For example, patient behavior, consumer or voter preferences, and epidemiological context are highly volatile.  In our empirical examples, we study a case where a policy change creates a rapid fluctuation in the cost of electricity and a case where changing epidemiological context and internet usage hamper the ability to predict flu trends.  In both cases, concept drift renders inference using these predicted outcomes invalid.    

The practitioner cannot eliminate uncertainty arising from concept drift without retraining the prediction model, which can cost upwards of tens of thousands of dollars or more and could be impractical in many settings. \citep{li2023flm101bopenllmtrain,sharir2020costtrainingnlpmodels}. The practitioner must therefore choose, using factors beyond the data, whether or not to trust the model.  The decision relies on the practitioner's \emph{preferences} which depends on the \emph{consequences} of making the wrong decision, which is inherently bespoke to a given situation. This could include the practitioner's tolerance for making a mistake (and the situation-specific consequences thereof), the cost of model fitting and data collection, and the ratio of signal to noise in the parameter of interest. 

In this work, we propose a pragmatic strategy for choosing when to refit complex prediction models, when the goal is {\it inference}, by exploiting a novel connection to portfolio optimization theory.
We compare three distinct strategies: (1) fully {\it refitting} the ML/AI model from scratch, (2) partially {\it recalibrating} predictions using a small new sample, or (3) simply {\it retaining} the existing model without update. The decision of whether to {\it refit}, {\it recalibrate}, or {\it retain} involves significant financial considerations. Fully {\it refitting} a large-scale predictive model can incur substantial computational resource, labor, and time costs~\citep{smith2023large}. {\it Recalibrating} uses a few labeled samples to adjust for the bias induced by the prediction model, although we do not know until after making the decision how effective this will be. We only know how well a similar procedure worked in a previous calibration period. In situations where the concept drift is rapid, the recalibration will be inferior to a full refit. {\it Retaining} the existing model without updates, while cost-free in the short term, risks a gradual decline in accuracy and relevance of the model over time \citep{vela2022temporal}. 

We approach {\it refitting} versus {\it recalibrating} using portfolio optimization theory, where this decision is treated as `investing' in one of two options, or `assets.' This powerful connection, to our knowledge, has not yet been exploited in the machine learning or statistics literature.  Portfolio theory comes with a natural language for expressing uncertainty and quantifying preferences.  {\it Refitting} the model has statistical uncertainty, or {\it risk}.  We know approximately how well our prediction algorithm will do based on prior fits of the model, but we expect some variation based on the inherent stochasticity of the prediction task. {\it Recalibrating} also introduces {\it ambiguity}, meaning that we do not know how much the prediction model has decayed from the last time it was trained and, thus, how effective our adjustment will be. Portfolio optimization theory has a similar conundrum, where an investor must choose between assets (stocks, bonds, etc.) that are volatile, but in a mostly predictable way, and more uncertain assets where the distribution of volatility is harder to predict. In both cases, it is impossible to characterize the optimal decision solely in terms of statistical properties -- the user has incomplete information in a stochastic environment, which necessitates introducing idiosyncratic user preferences to avoid either risk or ambiguity.

The third option, \textit{retaining},  we think of as the baseline choice. Retaining means that you believe the conditions that were used to fit your model now apply perfectly in your new setting (e.g. time or location). This is equivalent to assuming that your preferences are such that you have no aversion to ambiguity. As seen in our empirical application predicting flu trends, retaining your model is rarely reasonable in practice and can lead to misleading results.  Retain is preferred if the cost of collecting even a small amount of data to recalibrate is extremely high, and the cost of retraining a model is also high. In such cases, the decision-making process is not interesting since the practitioner has no choices to make. Retaining, therefore, is a possibility under some circumstances but not a typical part of our decision-making paradigm. We will illustrate examples of the difficulties of choosing between these strategies in two data-driven examples. One will illustrate the issues in elasticity and electricity price estimation, while the latter will reassess the infamous Google Flu algorithm and evaluate when the algorithm should have been retrained.


Our work is part of a newly flourishing literature that emphasizes the role of the modeler (and their preferences and context) in making consequential decisions that will impact the interpretation of statistical results. The idea that a data scientist should transparently and reproducibly report all steps in the discovery pipeline is a key idea in Bin Yu's ``Veridical Data Science'' \citep{Yu2020}. For example, quantifying the shared versus differing information learned from multiple models, and thus how likely one's ``best model'' may have been chosen by random chance, has driven the study of the ``Rashomon sets''\citep{ rudin2024amazingthingscomehaving,venkateswaran2024robustlyestimatingheterogeneityfactorial}.  Andrew Gelman brought to attention the connection between these decisions, which he calls ``The Garden of Forking Paths,'' to the subject of multiple comparisons and $p$-hacking \citep{Gelman2019TheGO}. Xiao-Li Meng described the effect of non-probabilistic decisions that one makes through his language on Design, Divine, and Device probability \citep{ddd}. In these and other works, much has been done to articulate, quantify, and describe how different choices at these decision points can lead to different models and potentially different conclusions. Our work, instead, focuses on systematically evaluating these options, weighing them against one's preferences, and developing rules to \textit{make} a decision in the face of this uncertainty.

\section{Framework}

\label{sec:problem}

\subsection{Modeling Assumptions and Goals}
\label{sec:setup}
\indent We are motivated by studies with an expensive or hard-to-obtain outcome, $Y$, and potentially associated covariates $X$. The relationship between $Y$ and $X$ is governed by a stochastic, data generation process, $G$, which changes with respect to $t \in [0, T]$. : Y(t) = G(X, t). The practitioner's goal is to describe the relationship between $Y$ and $X$ at time $t$ as parameterized by $\beta$ from interpretable model $\tilde{G}$. For instance, in political science, $Y$ could be a person's political affiliation or voting patterns, which may require costly in-person interviews or exit surveys, $X$ could be various demographic variables such as age, sex, race, $G$ would be the true psycho-social-economic process by which these factors influences a person's voting patterns, $\tilde{G}$, a logistic regression that a social scientist fits to model said the relationship, $\beta$, the coefficients of said logistic regression, and $t$ modeling how these preferences change over time. We use time to make our procedure concrete, but the method we propose is directly applicable in other continuous domains as well (e.g. distance).

\indent Further, the practitioner has access to a pre-trained AI/ML model, $\hat{G}_0$, which predicts $Y$ using $X$, though they may not have access to the operating characteristics of this model or the data used to train it. In the previous example, this could be a natural language processing model that predicts voting behavior from online footprint \citet{jiang2024donaldtrumpsvirtualpolls,10.1007/978-3-030-71704-9_17}. On one hand, such a model could potentially be a great boon. If the model provides reasonable estimates of $Y|X$ at a much cheaper cost, then one can eschew or cut down on the budget spent on collecting the hard-to-observe $Y$s. On the other hand, if the model is inaccurate or biased, then it is better to spend our budget mostly collecting real data and ignoring the predicted $Y$s. Further complicating the question, is the effect of concept drift. If the black-box model was fit in the past, then it is possible that it may have provided accurate estimates in the past, but is no longer useful for decisions in the future. In designing our decision-making framework, we would like our procedure to be able to take into account these various factors.

\indent At a high level, this leaves the practitioner with three potential strategies with regard to how to use the said model in their discovery pipeline, while balancing the cost to collect $Y$ versus $X$, the risk versus and the ambiguity:
\begin{enumerate}
    \item \textbf{Retain} the model - collect $X$, then use the previously trained model to impute $Y$ to get $\tilde{G}_0(X)$ and regress $\tilde{G}_0(X)$ on $X$ to estimate $\beta$. -  {\it Cheap} but {\it uncertain}. Responses may no longer be accurate/useful depending on model performance and the amount of concept drift.
    \item \textbf{Refit} the model- Collect new $(X,Y)$ data to refit $\tilde{G}_0(X)$ the proceed as Retain.-  {\it Expensive} but {\it certain}. Need to collect new training data, plus pay model fitting cost. Inherent/natural variation in data generating process (\textit{risk}).
    \item \textbf{Recalibrate} the model - {\it Intermediate} option. Use a few labeled cases to estimate the adjustment factor that accounts for concept drift. Natural variation in data generating process (\textit{risk}) and uncertainty in estimating the correction factor (\textit{ambiguity}). 
\end{enumerate}
In the succeeding sections, we will demonstrate how to build a decision-making framework for these three options and how to estimate the utility of each option.

\subsection{Making the Decision}
\label{sec:decision}


We approach the choice between {\it retaining}, {\it refitting}, or {\it recalibrating} as a decision theory problem where we are interested in making a decision in the presence of uncertainty. However, there are two sources of uncertainty: \textit{risk} and \textit{ambiguity}. \textit{Risk} refers to the effect that the inherent stochasticity of the data generation process has on parameter estimation. Due to the stochastic nature of the data generation process, one would expect to observe a different value for Y if the experiment were rerun. With a different $Y$, regardless of whether one chooses a retain, refit, or recalibrate approach, one would get a slightly different estimate for $\beta$. This is the most common form of uncertainty that traditional frequentist hypothesis testing is built to quantify.

On the other hand, \textit{ambiguity} refers to model uncertainty, or in other words, the uncertainty that arises from how uncertain one is about the true probabilistic model. For example, risk occurs when an actor is given perfect information about the data generation process $G$ but, due to natural stochasticity, chooses the wrong strategy. Ambiguity is when $G$ is not known and must be estimated, leading to a range of possible data generation models. Under this setup, decision-making is to be done without knowledge of the specification of the data-generating process or the AI/ML model, so both forms of uncertainty are present. In Section \ref{sec:estimating_lam}, we describe how to estimate both sources of uncertainty.

To build a decision framework that can handle risk and ambiguity, we employ a connection to a field that has well-studied the trade-offs of making decisions under different sources of uncertainty: portfolio optimization theory. Portfolio theory comes with a natural language for expressing uncertainty. One starts with a fixed budget for data collection. Then, one can treat choosing a strategy as `investing' in an option, and after pursuing the said option, one can observe whether it provided the best return, in our case, gave us the best estimate of our parameter of interest in terms of mean squared error (MSE). Moreover, this gives us the ability to easily include fixed costs, such as the cost of training an AI/ML model, which is often a substantial burden in many studies. The MSE here is for estimating the parameters of a downstream model we wish to draw inference on using outcomes predicted from our upstream AI/ML model and not the mean squared error of the AI/ML model itself.

To choose an appropriate utility function for the decision-making process, one needs a utility function that places a high value on obtaining the best-performing result while penalizing the uncertainty in said result. In portfolio allocation, this is oft described as the trade-off between the expected return of an option versus its volatility. We choose as our objective function the Arrow-Pratt utility function, modified to penalize risk and ambiguity \citet{maccheroni2013alpha}. We assume the person making the decision is a rational actor whose goal is to choose the strategy that maximizes an ambiguity-adjusted Arrow-Pratt utility function, $U(\cdot)$ of the form:
\begin{equation}
\label{eq:utility}
    U(MSE,\lambda,\theta) = -1\times E_P(MSE) - \frac{\lambda}{2} \sigma^2_P (MSE) - \frac{\theta}{2}\sigma^2_\mu (MSE).
\end{equation}
$E_P(MSE)$ is the expected MSE of our parameter of interest, $\beta$, $\sigma_P^2(MSE)$ and $\sigma_\mu^2(MSE)$ are the variation in the MSE that is attributable to sampling (ie, {\it risk}) versus variability in estimating the factor {\it recalibration} (ie, {\it ambiguity}), respectively. The parameters $\lambda$ and $\theta$ are positive real numbers that indicate how averse one is to risk and ambiguity, respectively, with higher values of $\lambda$ and $\theta$ corresponding to more aversion to risk and ambiguity, respectively. See Section \ref{sec:coef_choice} for a detailed discussion on how $\lambda$ and $\theta$ can be chosen in practice. Let $MSE_{rec}$, $MSE_{ref}$, and $MSE_{ret}$ be the MSEs of $\beta$ when taking each of three strategies of {\it recalibrating}, {\it refitting}, or {\it retaining}. Assuming that one does need to expend any additional funds when choosing to retain a model, the economic question becomes when given a fixed budget $c$, what is the optimal allocation of said budget to the recalibrate versus refitting strategies? Modeling $w_{rec}$ and $w_{ref}$ as weights such that $w_{rec} + w_{ref} = 1$, we seek to solve for the weights such that the combined asset:
\[
  w_{rec}\times MSE_{rec} + w_{ref}\times MSE_{ref}
\]
\noindent maximizes our utility function, Equation \ref{eq:utility}. Solving this yields a system of equations:
\[
\begin{aligned}
  \lambda& \begin{bmatrix}
    \sigma_{P}^2(MSE_{ref})  &  \sigma_{P}(MSE_{ref},MSE_{rec}) \\
    \sigma_{P}(MSE_{ref},MSE_{rec}) & \sigma_{P}^2(MSE_{rec})
  \end{bmatrix}
  \begin{bmatrix}
    \hat{w}_{ref} \\
    \hat{w}_{rec}
  \end{bmatrix} + \\[2ex]
  &\quad\quad \theta \begin{bmatrix}
    0 & 0 \\
    0 & \sigma^2 _\mu(MSE_{rec})
  \end{bmatrix} = 
  \begin{bmatrix}
    E_p(MSE_{ref}) - MSE_{ret)} \\
    E_p(MSE_{rec}) - MSE_{ret)} \\
  \end{bmatrix}.    
\end{aligned}
\]
\noindent where $\sigma_{P}(MSE_{ref},MSE_{rec})$ is the risk covariance for {\it recalibrate} versus {\it refit}. The solution occurs at:
\[
  \hat{w}_{rec} = \frac{BD - HA}{CD - H^2} \hspace{0.25in}  \hat{w}_{ref} = \frac{CA - HB}{CD - H^2},
\]
where
\[
\begin{aligned}
    A &= -1*(E_p(MSE_{rec}) - \overline{MSE_{ret}}) & \quad & D = \lambda \sigma_{P}(MSE_{rec}) + \theta \sigma_{\mu}(MSE_{ref}) \\
    B &= -1*(E_p(MSE_{ref}) - \overline{MSE_{ret}}) & \quad & H = \lambda \sigma_{P}(MSE_{ref}, MSE_{rec}). \\ 
    C &= \lambda \sigma_{P}(MSE_{ref}) \\
\end{aligned}
\]
Finally, while a fractional allocation between strategies is a valid approach when buying financial assets, for choosing between data pipelines, we must make a discrete choice, so we utilize the following strategy to discretize our options:
\begin{equation}
\label{eq:decision}
\text{Decision Rule} = 
\begin{cases}
    \text{ Refit is Optimal} &  \text{If } 
    \hat{w}_{ref} > \hat{w}_{rec} \text{ and } A >0   \\ 
    \text{ Recalibrate is Optimal} &  \text{If } \hat{w}_{ref} < \hat{w}_{rec} \text{ and } B>0  \\ 
    \text{ Retain is Optimal } &  \text{If } A <0 \text{ and } B<0
\end{cases}.
\end{equation}
In the decision rule above, for retaining to be the optimal decision, we require the expected MSE for recalibrating and refitting to be greater than retraining the model. This is highly unlikely as it would require the model refitting cost and the data collection cost to be high enough that a recalibration is worse than doing nothing. In such a scenario, the practitioner has practically no choices, so while retaining is included for completeness, we shall not see it chosen in the simulated or real data simulations below. In the Supplementary Material, we show the explicit mapping between this decision rule and the setup in the case of portolio optimization.




\subsubsection{Choosing $\lambda$ and $\theta$ }
\label{sec:coef_choice}
The above decision rule (\ref{eq:decision}) relies on user-specified preferences for risk ($\lambda$) and ambiguity ($\theta$). Soliciting such preferences is an active area of research, and there exists a large body of literature on the topic~\citep{Guidolin2012,Bhren2021}. Two promising approaches, as described in~\citet{https://doi.org/10.3982/QE1930}, are a {\it context-specific approach} versus a {\it hypothesis testing approach}. The context-specific approach elicits informative choices of $\lambda$ and $\theta$ using prior knowledge of the type of misspecification. The hypothesis testing approach is most salient when a user lacks such domain knowledge. 

We develop a strategy based on the hypothesis testing approach.  Immediately after AI/ML model training, we expect minimal drift, and thus {\it retaining} is likely the best option. As we are dealing with a stochastic decision-making process, there is always a nonzero probability that we could choose to {\it refit} or {\it recalibrate} during this period. We can decrease the probability that {\it refitting} or {\it recalibrating} is chosen by increasing $\lambda$ and $\theta$, our aversion parameters. We choose values of $\lambda$ and $\theta$ such that, early on, we have a less than $\alpha$ probability of choosing anything but to {\it retain} the current model by setting:
\[
  \hat{\lambda} = \arg\min_{\lambda >0 }\Pr[U(MSE_{ref},\lambda,0) > U(MSE_{ret},0,0)] \leq    \alpha.
\]
Then, using $\hat{\lambda}$ as a plug-in estimator, we choose $\theta$ such that
\[
  \hat{\theta} = \arg\min_{\theta >0 }\Pr[U(MSE_{rec},\hat{\lambda},\theta) > U(MSE_{ret},0,0)] \leq \alpha.
\]  
This gives a simple approach for choosing $\lambda$ and $\theta$ that controls the Type I error rate under a null hypothesis that there is no drift. Alternative approaches based on likelihood-ratio testing~\citep{https://doi.org/10.3982/QE1930}, robust control theory~\citep{Ioannis}, and robust sequential hypothesis testing~\citep{5070f351-a6e6-39ad-a9fd-56d395379352} could also be used.

\subsection{Estimating the Mean Squared Errors of Each Strategy}
\label{sec:estimating_lam}

To employ the above decision rule, one needs a procedure to estimate the amount of risk and ambiguity that arises when estimating $\beta$ across each of the three strategies. To do this, we shall say that the parameter of interest $\beta$ comes from the minimization of the convex objective function:
\[
  \hat{\beta} = \arg\min_{\beta} h\left(\hat{Y}, X;\beta\right).
\]
In the succeeding sections, we shall assume that $h$ is the objective for a generalized linear model, however much of the intuition will hold for other convex estimation problems. Let $\hat{Y}_{ret}$, $\hat{Y}_{ref}$, $\hat{Y}_{rec}$, and $Y$ be the predicted $Y$ under the {\it retain}, {\it refit}, and {\it recalibrate} strategies, and true $Y$, respectively. Further, let $\hat{\beta}_{ret}$, $\hat{\beta}_{ref}$, $\hat{\beta}_{rec}$, and ${\beta}$ be the resulting parameter estimating resulting from the 3 strategies as well as the true $\beta$.

Since the MSE depends on the number of labeled/unlabeled observations we collect, we also need to specify the relationship between budget and data collection. Let our total budget be $c$,  let model refitting cost be $c_{model}$, let collecting $(X, Y)$ together cost $c_{labeled}$, and let collecting $X$ alone cost $c_{unlabeled}$. If one retains a model, no model refitting cost is needed, and one can use the entire budget to collect $\frac{c}{c_{unlabeled}}$ samples to estimate $\beta$. If one refits a model, one must first spend $c_{model}$ for refitting, leaving a budget of $\frac{c - c_{model}}{c_{unlabeled}}$ samples. Finally, if one chooses to recalibrate, choose the optimal ratio of labeled to unlabeled data ( costing $\frac{c}{c_{labeled}}$ and $\frac{c}{c_{unlabeled}}$) respectively to best estimate $\beta$.

\subsubsection{Strategy 1: Retain the Model}
The simplest strategy is to use the original AI/ML model for everything. In such a case, we would expect the bias of $\hat{\beta}_{ret}$ to grow with increasing distance from the initial model fit (e.g. time, space, or distance). To measure the rate at which this bias grows, we shall assume that post-initial training, the maintainers have collected data at regular time intervals to evaluate the performance of the model post-deployment. This is referred to as the monitoring, maintenance, verification, or, the moniker we shall use, the calibration phase. During the calibration phase, we shall repeatedly compare $\beta$ and $\hat{\beta}_{ret}$ and fit a time series model to estimate the expected MSE at $t_{future}$. This allows us to extrapolate out how well retaining the model works from the calibration data, where we observe data, to $t_{future}$ where we are yet to observe data and a decision needs to be made. This can be done via the algorithm below:

\textbf{Retain: Do-Nothing Algorithm}

\begin{enumerate}
    \item For each calibration point $t$ in calibration data $(\tau_1,...,\tau_{T})$,
    \item[] Compute $\hat{\beta}_{ret,t} = \arg\min_\beta g(Y_t, X_t; \beta)$
    \begin{enumerate}
        \item For $j$ in $1, ... , B$, take a bootstrap sample $(X_{t}^b Y_{t}^b)$ of size $c/c_{unlab}$
        \item Compute $\hat{\beta}_{ret}^b =  \arg\min_\beta g(Y_t ^b, X_t ^b; \beta)$
    \end{enumerate}
    \item  $\hat{r}_t = ||\hat{\beta}_{ret,t} - Median(\hat{\beta}_{ret}^1, ... ,\hat{\beta}^B)||_2^2 $
    \item Fit a time series on  $(\hat{r}_{\tau_1}, ... ,\hat{r}_{\tau_T})$.
    \item $MSE_{ret}$ be the predicted value at $t_{future}$ using said time series
\end{enumerate}

The bootstrap sample size of $c/c_{unlab}$ illustrates the fact that since {\it retaining} does not require model {\it refitting} or additional labeled data, we get our best estimate for the MSE by spending it all on the unlabeled data, $X$, predicting $Y$ using our retained AI/ML model, and computing its MSE.

\subsubsection{Strategy 2: Refit the Model}

If AI/ML model performance has degraded considerably, the best option may be to retire the old model and {\it refit} a new one. 
We make the assumption that any newly fit AI/ML model, when trained on the new data, will have similar accuracy to the previous model. To refit, one must spend $c_{model}$ of their budget to train a new model and the remaining  $(c - c_{model})/c_{unlab}$ for unlabeled samples to estimate the accuracy of the estimates using the new model. If we allow $h$ to be a generalized linear model, we have the traditional decomposition \citet{agresti2015foundations}:
\[
  MSE(\hat{\beta}_{ref}) = Var(\hat{\beta}_{ref}) = (X^T W X)^{-1}.
\]
Here, $(X^T W X)$ is the information matrix for $\beta$, $W$ is the corresponding weight matrix and  $(X^T \hat{W} X)$ is the asymptotic covariance estimated at $\hat{\beta}$. We estimate $MSE(\hat{\beta}_{refit})$ using the bootstrap procedure described below:\\
\textbf{Refit Algorithm}
\begin{enumerate}
\item For b in 1, ..., B
    \begin{enumerate}
    \item Take a bootstrap $(X^b, Y^b)$ of size $(c - c_{model})/(c_{unlab})$ from the data that was held out during initial model training 
    \item Compute $\hat{Var}(\hat{\beta}_{ref}^b) = (X^{bT} \hat{W}^b X^b)^{-1}$ where  $\hat{\beta}_{ref}^b =  \arg\min_\beta g(Y_t ^b, X_t ^b; \beta)$
\end{enumerate}
\item $MSE_{ref}  = $Median($\hat{Var}(\hat{\beta}^1 _{ref}), ..., \hat{Var}(\hat{\beta}^b _{ref}))$
\end{enumerate}

\subsubsection{Strategy 3: Recalibrate the Model}

The middle ground between {\it retaining} and {\it refitting} is to {\it recalibrate} the model. This requires us to first spend our budget on collecting an appropriate ratio of unlabeled and labeled data.
Regressing the AI/ML-predicted outcomes from the unlabeled data against $X$ yields a biased estimate of the parameter of interest. Using the labeled data, we compare our AI/ML generated $\hat{Y}$ to real $Y$ to learn the appropriate correction factor. Combining the biased parameter estimate with its appropriate correction factor to conduct valid statistical inference is what \citet{hoffman2024really} call ``Inference on Predicted Data.'' Here, we utilize a mature procedure for such inference known as ``\texttt{PPI++}'' \citep{angelopoulos2023prediction,angelopoulos2024ppi}, but other approaches that yield valid parameter estimates will work as well. Given the associated loss function $l(X,Y; \beta)$ for estimator of $\beta$, \texttt{PPI++} is a parameter estimation procedure that minimizes:
\begin{align*}
    \hat{\beta}_{rec} &= \arg\min_{\beta} l(X_{unlab,i}, \hat{Y}_{unlab,i};\beta)  + \gamma * (l(X_{lab,i}, \hat{Y}_{lab,i};\beta)  - l(X_{lab,i}, Y_{lab,i};\beta) )  \\
    &= \hat{\beta}_{ref} - \gamma * \hat{\Delta}
\end{align*}
where $\gamma \in [0,1]$ is a data-estimated tuning parameter and $\hat{\Delta}$ is the "Rectifier" as defined in \citet{angelopoulos2024ppi}. When $\gamma $ is close to 0, this indicates that the AI model perfectly captures the Y|X relationship, and the parameter estimates require no correction. Using the unbiasedness of the Prediction-Powered Inference estimator \citep{angelopoulos2024ppi}, we can decompose the MSE via:
\[
\begin{aligned}
  MSE(\hat{\beta}_{rec}) &= Var(\hat{\beta}_{rec}) \\
  &= Var(\hat{\beta}_{ref}) + \gamma^2 Var(\hat{\Delta}_{ref})\\
  &= (X_{unlab} ^T \hat{W}  X_{unlab})^{-1} + \gamma^2 (X_{lab} ^T \hat{D}_{ref} X_{lab})^{-1}
\end{aligned}
\]
where $(X_{lab} ^T \hat{D}_{ref} X_{lab})$ is the information matrix for $\hat{\Delta}$. Note that $Var(\hat{\beta}_{ref})$ describes the amount of risk in the Prediction-Powered Inference estimate of $\beta$ while $Var(\hat{\Delta}_{ref})$ describes the amount of ambiguity. To extend this recalibration procedure to our temporal setting, we make the following extension to the PPI++ procedure: 

\vspace{2ex}

\textbf{Recalibrate Algorithm}

\begin{enumerate}
    \item For each calibration point $t$, perform PPI++ and compute $\hat{W}_{i, ref}, \hat{D}_{i,ref},$ and $ \hat{\gamma}_i$.
    \item Fit time series for $\hat{W}_{i, ref}, \hat{D}_{i,ref},$ and $ \hat{\gamma}_i$.
    \item Using said time series, estimate the predicted distribution for $\hat{W}_{i, refit}, \hat{\Delta}_{i,ref},$ and $ \hat{\gamma}_i$ at $t_{future}$.
    \item For each sample size ratio $\zeta \in (0, ... , 1)$:
    \begin{enumerate}
           \item For $b = 1, ... B$:
        \begin{enumerate}
            \item Draw $\hat{\gamma}^b$, $\hat{W}_{i,ref}$, and $\hat{D}_{i,ref}$ from their forecast distributions.
            \item Draw $(1-\zeta) * \frac{c}{c_{unlab}}$ samples of $X_{unlab}$ and  $(\zeta * \frac{c}{c_{lab}}$ samples of $X_{lab}$ with replacement. Call these $X_{unlab}^b$ and $X_{lab}^b$ respectively.
            \item Let $MSE_{ref}^b(\eta) = (X_{unlab} ^{bT} \hat{W}^b  X_{unlab})^{-1} + \gamma^{b2} (X_{lab} ^{bT} \hat{D}_{refit}^b X_{lab}^b)^{-1} $ .
        \end{enumerate}
        \item Let $MSE_{rec}(\eta) = Median(MSE_{rec}^1(\eta), ... ,MSE_{rec}^B(\eta))$.
    \end{enumerate}
    \item $MSE_{rec} = \min_{\eta}(MSE_{rec}(\eta))$
\end{enumerate}

\noindent As time increases, we expect the original parameter estimates to become increasingly unreliable. Thus, $\hat{\gamma}$ and $\hat{W}$ should decrease, while $\hat{\Delta}$ should increase. 



\section{Simulation Study}
\label{sec:sims}
We start by showing a simple simulation in which the ground truth is known. We generate 10,000 samples of $X_1, X_2$, independent standard normal variables. We then generate
\[
  Y(t) = X_1\cdot \exp(0.25t) +  10\cdot I(X_2 >= 0) + \epsilon,
\]
where $I(\cdot)$ is the indicator function, and $\epsilon \sim N(0, 0.05)$. We start with a model that is trained to predict $Y$ from $X_1$ and $X_2$ at time $t = 0$. Then a downstream model estimates the relationship between $X_1$ and $Y$ via a linear regression, $
  \hat{Y} = X_1 \hat{\beta} + \hat{\epsilon}$.
If the prediction model is accurate at time $t$ (or location/distance t), $\hat{\beta}$ should be close to $\exp(0.25t)$. As $t$ increases, the prediction model will be less accurate, and our model evaluation strategy will change from {\it recalibrate} to {\it refit}. As our data-generating mechanism contains a step function, a reasonable choice for the prediction model is to use random forests trained with 5-fold cross-validation.

We simulate a calibration period between times $t = 0$ and $t = 3$, where we have 50 distinct points in which 200 labeled and unlabeled samples were collected. We seek the decision to {\it refit}, {\it recalibrate}, or {\it retain} at time point $t = 3.5$. Our total budget to make a decision is 200, the cost of model-fitting was 10, unlabeled data, $(X_{1,unlab}, X_{2,unlab})$, each cost 1 to collect, and labeled data, $(X_{1,lab}, X_{2,lab}, Y_{lab})$, cost 2. As we are interested in a parameter from linear regression, $\hat{W}$ and $\hat{D}_{refit}$ has a simple representation:
\[
MSE(\hat{\beta}_{rec})
    \approx \hat{\sigma}_{unlab} (X_{unlab} ^T  X_{unlab})^{-1}+ \gamma^2 * \hat{\sigma}_{unlab} (X_{lab} ^T  X_{lab})^{-1}
\]
and $X = [X_1, X_2]$. The approximation here comes from how the classical $\hat{\sigma}_{unlab} (X_{unlab} ^T  X_{unlab})^{-1}$ estimator for the variance-covariance matrix is known to be biased when $Y$ is nonlinear. The common solution to this is to employ the sandwich estimator \citep{5805f73c-4dfa-385e-bd6d-68424fb9f5be, f457a0f7-5c1e-3d02-9161-7806a34faad9, c22b54d7-c957-32d0-8c11-13c0d22364bc}, but in practice, we find the extra number of parameters this requires makes it ill-suited for the extrapolation we require.


\begin{table}
    \centering
    \caption{Mean MSE and Variances for the 3 options from the Simulation Study }
    \begin{tabular}{|c|c|c|c|}
\hline 
 & Mean MSE & Variance from Uncertainty &  Variance from Ambiguity \\
\hline
     Retain & 0.133 & -- & --  \\ \hline
    Recalibrate & 0.092 & 5.55e-06 & 8.28e-06 \\\hline
    Refit & 0.073 & 9.10e-05 & -- \\ \hline
\end{tabular}
    \label{tab:simtab}
\end{table}

\begin{figure}[ht]
    \centering
    \includegraphics[width = 1\textwidth]{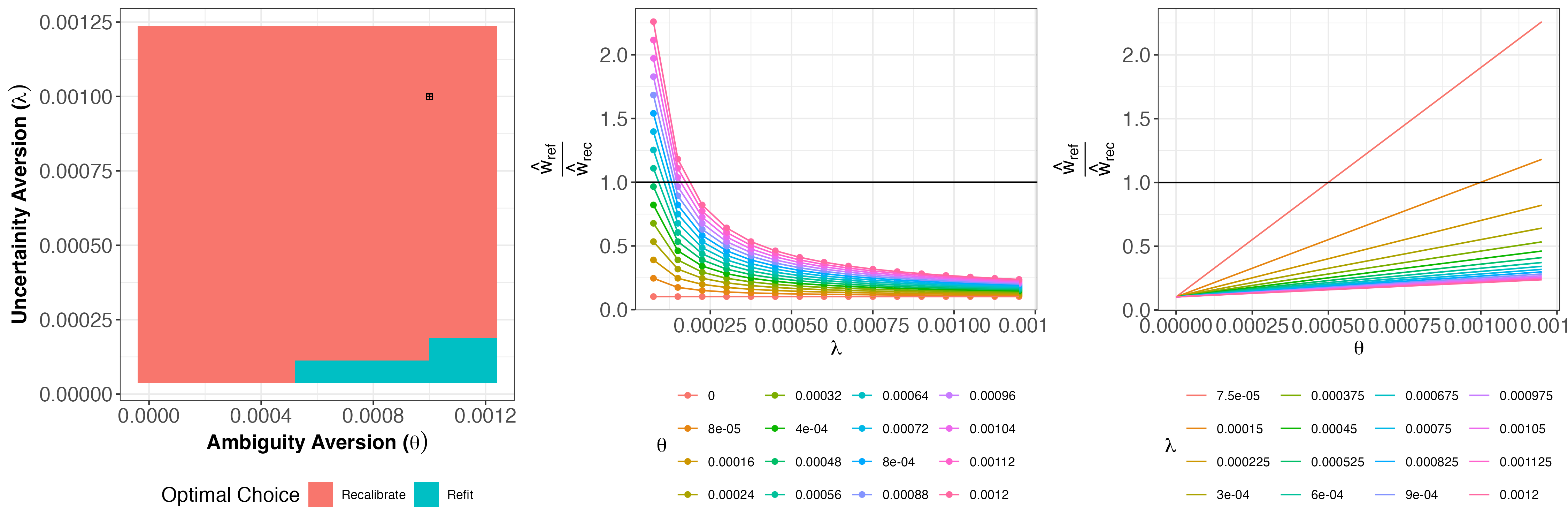}
    \caption{(Left): Our estimate of the best decision for various levels of aversion to uncertainty ($\theta$) and ambiguity ($\lambda$). The hypothesis-driven choice for $\lambda$, $\theta$ in Section~\ref{sec:coef_choice} is marked with a square. (Middle): $\frac{\hat{w}_{ref}}{w_{rec} }$  and $\lambda$ for various levels of $\theta$. Values above 1 choose refit as optimal. Above the horizontal line denotes {\it refit}, below is {\it recalibrate}. (Right): $\frac{\hat{w}_{ref}}{w_{rec} }$  for various $\theta$'s and $\lambda$'s. Above the horizontal line is {\it refit}, below is {\it recalibrate}. \label{fig:sim} }
\end{figure}

With the settings above, each MSE estimation algorithm was run with 100 bootstrap samples on an M2 Max Macbook Pro with 32 GB of RAM, giving us the mean MSEs and variances as seen in Table \ref{tab:simtab}. From this, we see that the {\it retain} strategy has the worst average MSE, followed by {\it recalibrate} and then {\it refit}. However, the {\it recalibrate} MSE has lower variance than the {\it refits} MSE, so our preference between the two is determined by how averse we are to ambiguity. In Figure~\ref{fig:sim}, we evaluate our decision rules with $\lambda$ varying from 0 to 0.00125 and $\theta$ varying from 0 to 0.0012. We find a clear trade-off where one prefers to {\it recalibrate} when uncertainty aversion is high and {\it refit} when ambiguity aversion is high. The utility function chosen by our Type I error rate control with $\alpha=0.05$ indicates a preference for {\it recalibration} (Figure \ref{fig:sim}, left panel, black cross). At no point is {\it retention} the optimal choice under these simulation settings.

\section{Electricity Forecasting in New South Wales}
\label{sec:electric}

As an example of decision-making using this procedure, we put ourselves in the situation of a large data center in New South Wales, Australia that requires 2000MWh of electricity~\citep{Afl_2023}. In early 2024, this is a little over half of the capacity of the largest data center in the world ~\citep{data_center_power}. To estimate electrical costs, the company created a two-step model forecasting electrical costs. First, every half hour, demand was forecasted for electrical demand  $\hat{e}$ using data from the previous 7 days. Second, forecasts for the day are regressed against price to obtain and estimate the daily price elasticity, $\beta$ from the equation: $\hat{e} = \beta X_{price} + \hat{\epsilon}. $All estimation was done using the Elec2 dataset~\citep{nsw} downloaded from Kaggle~\citep{Sharan_2020}. The data were pre-scaled so traditional log-log transformation was omitted~\citep{0f8094d1-2ebf-3360-90f1-98e50c172f31}. The reliability of this procedure is based on the AI/ML demand model, which provides accurate forecasts of electrical demand. This makes this procedure sensitive to unpredictable shocks such as natural disasters or system failures. As an example of such a shock, midway through 1997, the electrical grid of New South Wales joined the grid in the neighboring state of Victoria to allow for load sharing. Grid unification leads to changes in both the total supply and total demand so is an example of concept drift changing the relationship between the elasticity (the outcome) and the time of day (the covariate of interest). The decision maker, of course, may not always know when a shock (or gradual drift) occurs. We use a calibration period of 180 days with 60 days before grid unification and 120 days after. As we see the ground truth, we expect {\it recalibrating} to be appealing at first and then {\it refitting} to be preferable after unification. The decision-making process does not rely on knowing when, or if, unification has occurred {\it a priori}. 

Our demand forecasting model is a long short-term memory model (LSTM) of 336 cells (equivalent to one week of demand data measured every 30 minutes) implemented using the Keras machine learning library. The model was trained on 11,150 data points at the start of the study. The LSTM had a $\tanh$ activation function, a sigmoid function for the recurrent activation function, and no dropout. All other parameters were set to the default as determined by the \texttt{keras3} package in R.

\begin{figure}[ht]
    \centering
    \includegraphics[width = 0.45\textwidth]{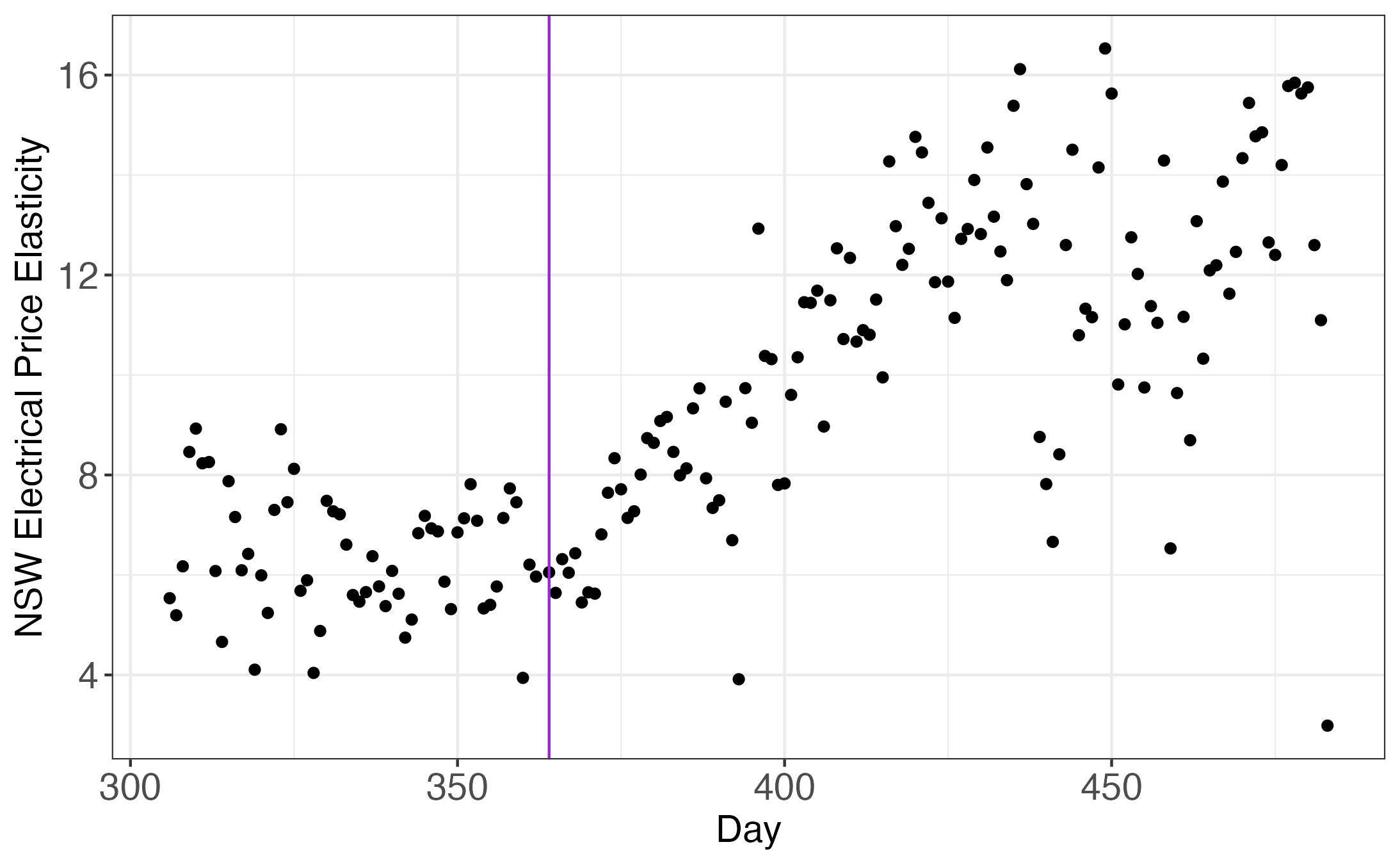}
    \includegraphics[width = 0.45\textwidth]{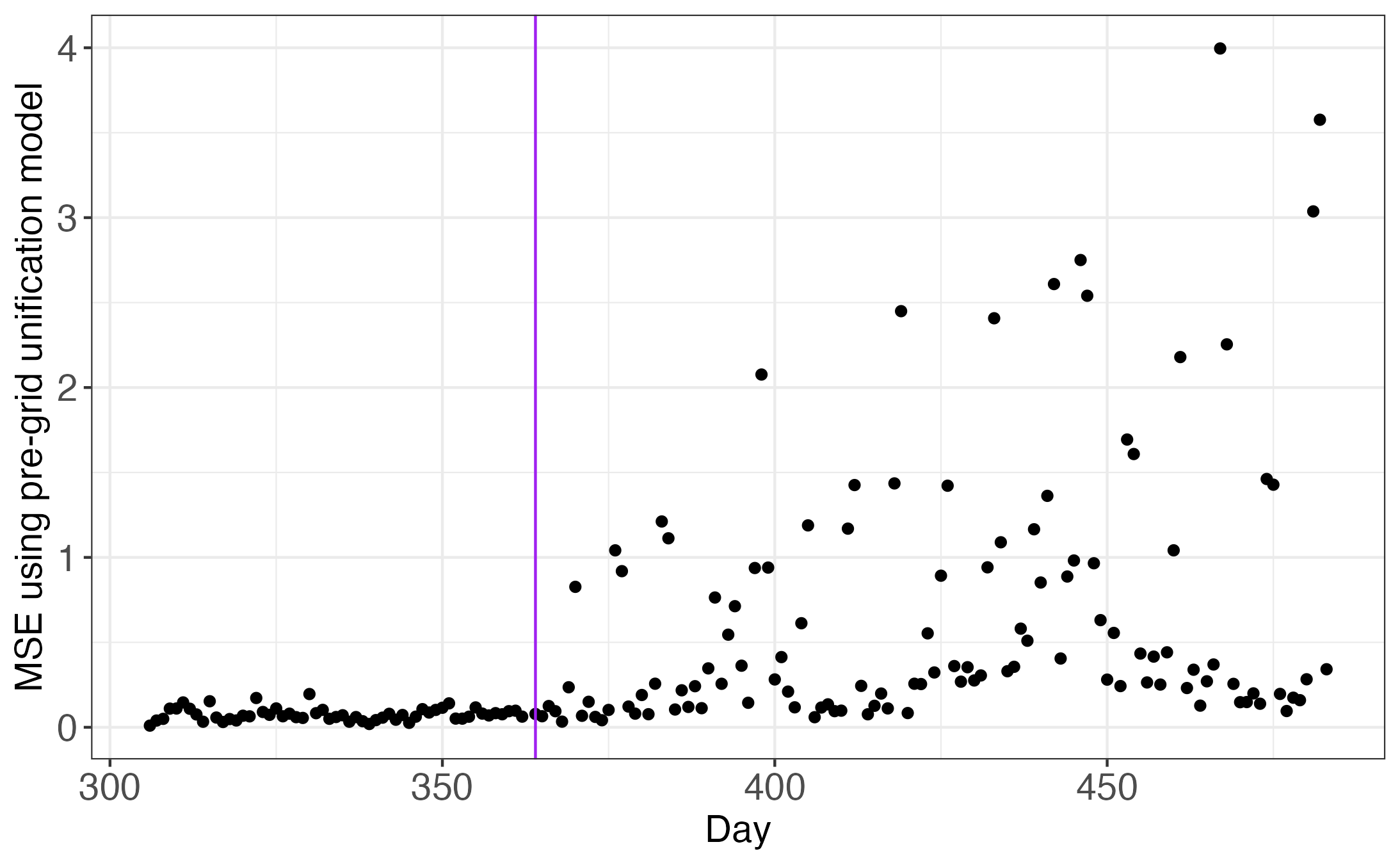}
    \caption{(Left): Electrical price elasticity over time. (Right): MSE of estimated electrical price elasticity using the pre-unification forecast model. The vertical purple line represents the grid unification day. The elasticity and our ability to estimate rapidly decay, signifying a case of concept drift.}
    \label{fig:NSW_change}
\end{figure}

We can see from Figure \ref{fig:NSW_change} that as soon as the grids were unified, the price elasticity nearly doubled from 6 to 15. Consequently, our errors in estimating $\beta$ increase 500\%, from 0.186 to 0.903. Each day, the model makes 48, 30-minute-ahead demand forecasts. Our company would like us to decide using this data and a budget of \$4,800. For costs, the cost of fitting the model was set at \$ 4,000 and our local electrical provider has offered to provide us \$120 for each instantaneous measurement of demand and price $(c_{labeled})$ or \$100 for just the price $(c_{unlabeled})$.


\vspace{-2ex}

\begin{figure}[ht]
    \centering
    \includegraphics[width = \textwidth]{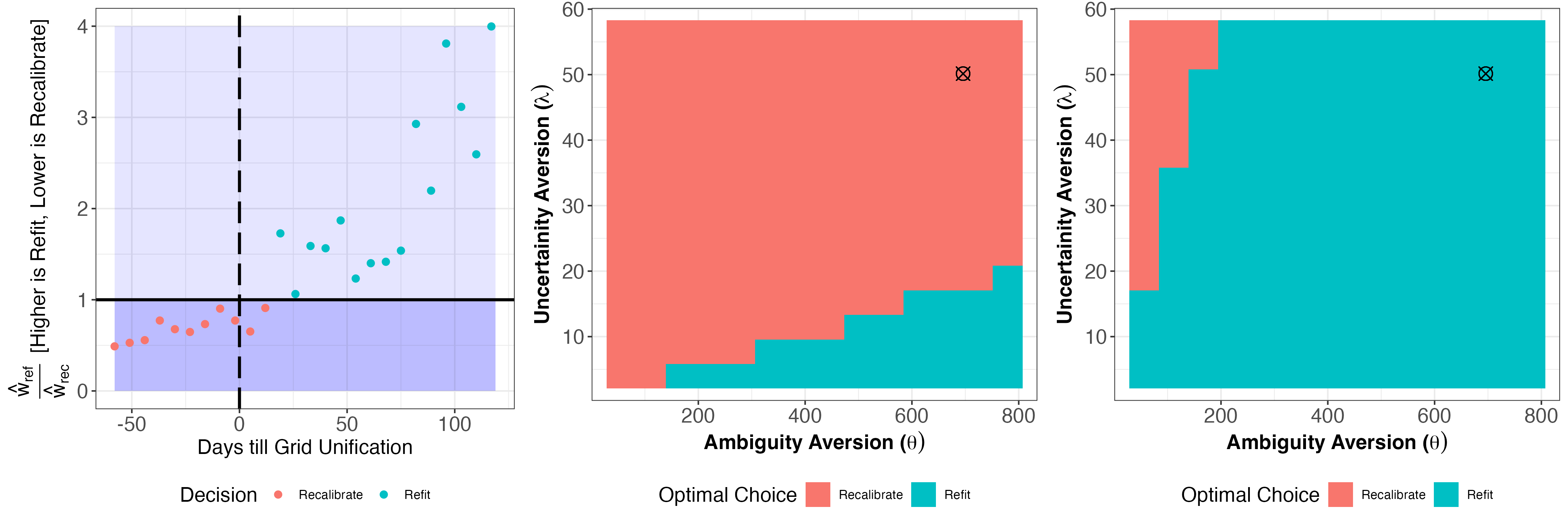}
    \caption{(Left): The decision that we chose using our Type I error rate controlled utility function at each week up to and after grid unification. (Center): Choice preference 60 days \textit{before} grid unification for a variety of utility functions. The black cross represents our Type I error rate controlled procedure. (Right): Choice preference 120 days \textit{after} grid unification for a variety of utility functions. }
    \label{fig:NSW_change2}
\end{figure}

\begin{table}[!ht]
    \centering
    \caption{True values for mean MSE of $\beta$, estimated electrical bill for one month running the data center, and the sum of differences between what we expected to pay versus what we paid }
    \begin{tabular}{|c|c|c|c|c|c|}
    \hline
         &Est.  Avg MSE & True Avg MSE  & Est. Electrical Bill 
 & Total Error  \\ 
        \hline
        Retain &1.52 & 1.37&  \$ 34,629,695  &  \$ 1,263,295 (\$ 809) \\ \hline
        Recalibrate &1.14 & 0.73 & \$ 32,853,507 & \$ 837,450 (\$ 1,540) \\\hline
        Refit& 0.123 & 0.08 & \$ 33,942,799 & \$ 411,145 (\$434)  \\ \hline
    \end{tabular}
\end{table}

We see in Figure \ref{fig:NSW_change2} that while our preference for {\it refitting} is steadily increasing as we move away from the calibration period, this rate of change rapidly increases after grid unification. This is echoed in our Type I error rate-controlled utility function which chooses to {\it recalibrate} before grid unification and {\it refit} afterward. {\it retain} was not optimal at any point.  

To validate the accuracy of our decision,
we decided to take all three options and computed the resulting $\beta$ estimates for a month, as well as our accuracy in estimating electrical costs. As we had predicted, {\it refit} was the most accurate, followed by {\it recalibrate} and {\it retain}. Moreover, of a total electrical bill of \$ 33,307,959, the {\it retain} and {\it recalibrate strategies} were off by roughly a million dollars, while {\it refit} was only off by around \$400,000. This shows that we are able to identify the most optimal of the three options.

\section{Google Flu Trends}
\label{sec:GFT}
Google Flu Trends was a high-profile initiative launched in 2008 that used large-scale tracking of Google search queries about illness-related terms as a low-cost way to estimate the prevalence of influenza-like illness in the United States \citep{Ginsberg2009}. It initially launched to great excitement due to its impressive predictive accuracy during the 2003-2008 flu seasons as well as its cheaper costs compared to the existing national-wide flu monitoring system, the Center for Disease Control and Prevention's Influenza Hospitalization Surveillance Network. However, after a series of high-profile failures, such as overestimation of twice the true number of flu-like illnesses during the 2012 and 2013 season \citep{155056}, and demonstrations that Google's predictions did not consistently outperform a simple forward projection of 2-week-old CDC data \citep{simple}, this lead to the project's abandonment in 2015.

While many analyses have been done on the project's failure, a consistent theme was how static Google's forecasting model was \citep{155056,Butler2013,wiredWhatLearn,Franzn2023}. Unlike models whose parameters can dynamically change over time as the diseases ebb and flow, the Google Flu Trends model was fit on weekly CDC data from 2003-2008 and minus an ill-documented update in 2009, was never refit or retrained till abandonment. This made the model ill-suited to new events such as the off-season 2009 spring epidemic \citep{Schmidt2019}, media-stoaked panic about the flu during the 2013 season \citep{155056}, and changes in Google's search algorithm (of which the official Google search blog reported 86
changes in their search algorithm from June and July 2012 alone \citep{155056}). This makes this a mix of both gradual and fast concept drifts as the relationship between the ILI counts and the covariates of epidemiological interest (in this case geographic region).

For this reason, Google Flu Trends serves as a good example of the importance of making sure an algorithm is kept up-to-date. To this end, we would like to recreate the problem these Google Researchers face: ``How often should we be refitting the Google Flu model to new data?'' and demonstrate how our procedure could have been used to suggest a refitting time before the catastrophically wrong predictions during the 2012-2013 seasons occurred. 

We obtained weekly data of the CDC and Google estimated number of influenza-like illness (ILI) cases from 2003-2015 for each of the 10 Health and Human Services Regions using the \texttt{epidatr} R package from the Delphi project \citep{delphi}. While changes in methodology since 2015 meant that the CDC's numbers now differ substantially from Google's past estimates, we found that taking a region-specific linear transformation of Google's estimates gave us a correlation between predicted and true ILI counts that matched those of the original paper (range: 0.79-0.9 vs 0.80-0.96). The data was then divided into three time periods to mirror the information the Google Flu team had access to. Data from 2003-2006 were treated as training data. We considered 3 different lengths of calibration periods 1) January 1, 2006-- June 30, 2008 (soon after the training period), 2) January 1, 2006 -- June 30, 2009  (in between), and 3) January 1, 2008 -- June 30, 2011 (right before big misses of 2012-2013). Budget costs for this procedure are difficult to determine as Google did not release cost estimates for their model, so to be conservative, we will have CDC data only cost twice as much as 
Google data to produce. We will thus set a total budget of 200, the model training cost was set to 10, and unlabeled and labeled data were given costs of 1 and 2 respectively. The upstream, black-box model was Google Flu Trend's ILI estimates with our region-specific recalibration. The regression of scientific interest is ILI regressed on region, specifically the regression coefficient for Region 2 which covers New York, New Jersey, Puerto Rico, and the Virgin Islands. Thus, each procedure evaluates if one is interested in obtaining the best estimates of ILI counts for the upcoming flu season in Region 2, one should retain, refit, or recalibrate their model. 

\begin{figure}[ht]
    \centering
    \includegraphics[width=1\linewidth]{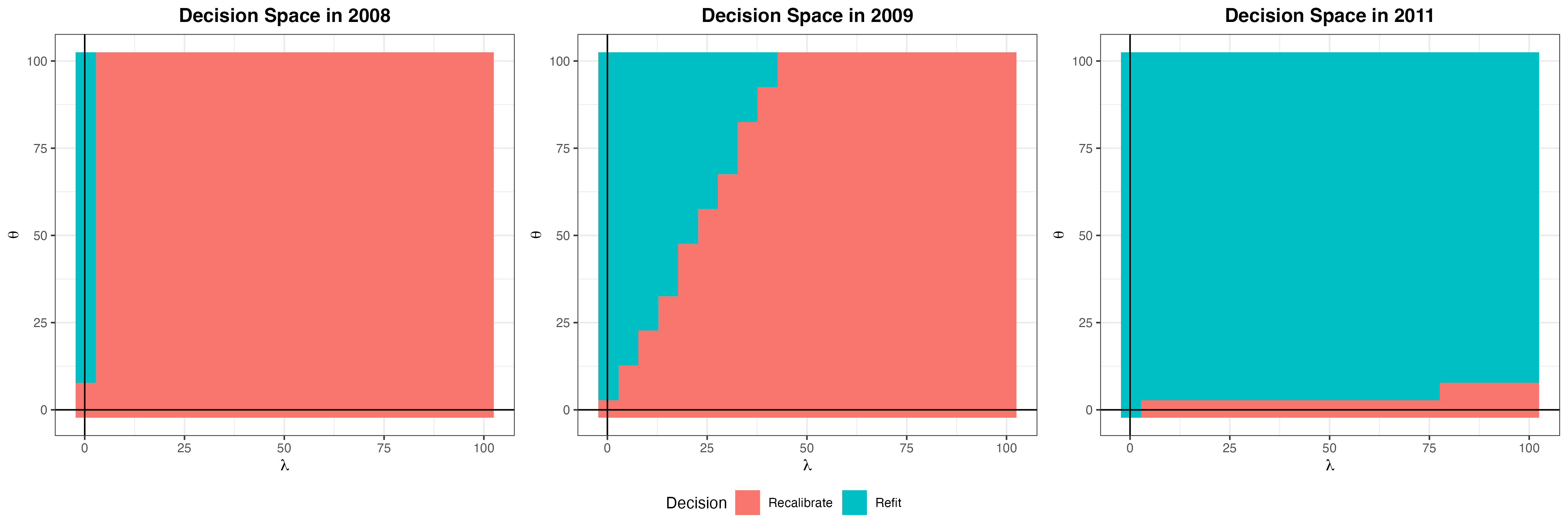}
    \caption{Decision space for the Refit vs Recalibrate for Google Flu Trends in 2008 (left) 2009 (middle) and 2011 (right) across different uncertainty and ambiguity aversion parameters.}
    \label{fig:flu}
\end{figure}
Figure \ref{fig:flu} shows the decision space for our analysis of the 2008, 2009, and 2011 flu seasons. In 2008 we can see that the decision space is almost completely in favor of keeping the model with only those with no aversion to ambiguity favoring a refit. In 2009, we now see an equal division of the space between those who favor refit versus those who believe a recalibration is a better approach. This increase is indicative of the model getting more and more out of date. Finally, we see by 2011, only one year before the major failures of the 2012-2013 season, almost all the results were in favor of refitting the model. Taken together, our analysis corroborates the existing findings that a decrease in model accuracy over time played a major role in Google Flu Trends missing the 2011-2012 seasons. While determining which exact aversion to uncertainty and ambiguity would work best for Google requires knowledge of Google's risk preferences that we are not privacy to, across a large range of preferences, it seems that Google Flu Trends would have been best served by \textit{refitting their model sometime between 2008-2009} and definitely before 2011. Moreover, this framework was not only able to give an explicit choice between strategies for maintaining the Google Flu model but was able to make this decision using the \textit{information that was available} at the time, not as a part of a retrospective assessment.

\section{Discussion}
\label{sec:discussion}
As complex prediction models play ever more essential roles across nearly every sphere of inquiry, it has become more and more pertinent to answer the question: how much should one rely on outcomes predicted by an AI/ML model? By connecting this to a deep literature on portfolio optimization, we suggest a new framework for settings where the goal is statistical inference.  After all, we cannot expect statistical uncertainty to vanish just because outcomes are predicted with sophisticated models. 
We show that with this toolkit, one can create actionable decision rules based on your budget, the cost of model training, the cost of data, and the performance of the models, which can accurately determine the most efficient way to keep relevant AI models up-to-date. We show this in the context of downstream statistical inference, but the asset allocation framework has the potential to be extended to many other metrics used to evaluate AI/ML quality such as accuracy, fairness, and interoperability.

Though this framework was designed with modern data science problems in mind, many of the topics connect to older discussions in statistics. Evaluating whether one's resources are best spent collecting cheaper imputed or more expensive real outcomes has long been known in the sampling literature under the names of ``two-phase sampling'' and ``double sampling'' \citep{be757a2c-b4ab-3579-8056-ad44dea4fec1, two_phase_1,bose}. However, much of this literature is developed in situations that require the imputation procedure or the downstream inference model to be well specified \citep{Cochran1977-py,Fujii2019,Davidov2000OptimalDF}. However, such assumptions are difficult to verify when the purpose of many black-box procedures is that they model things that are difficult to interpret. Thus, newer approaches that will give unbiased estimates regardless of the specification of the true data generation or black-box procedure such as \texttt{PPI++} are preferable.


There are two important limitations of note. First, we assume that if a model were to be retrained in the future, one would get an MSE distribution that resembles the one during initial model training. This makes this procedure ill-suited when the signal-to-noise ratio in the data changes due to external factors influencing the ``learnability'' of the outcome. The second assumption is in the choice of how to forecast $\hat{W}$ and $\hat{D}$. 
We forecast both $\hat{W}$ and $\hat{D}$ using simple linear regression. In cases where more data is available, more complex forecasting methods can be used.

As researchers integrate AI technology across all levels, from domains where data collection is cheap and plentiful (such as web-based text or image data) to domains where data are siloed and expensive (such as healthcare), our framework provides a new language to evaluate how to spend limited resources. This can be helpful in situations such as clinical trials, where there is the potential to enrich human subjects with AI-approximated data points and save resources while meeting traditional efficacy and safety requirements.

\newpage

\bibliographystyle{plainnat}
\bibliography{biblio}

\newpage

\end{document}


\section*{Supplementary material}
\renewcommand{\thesubsection}{S\arabic{subsection}}

\subsection{Portfolio-AI/ML Decision Equivalence}
\label{sec:equiv}
Here we describe how our maximizing the portfolio optimization utility function is equivalent to choosing the optimal utility for the ML setting. Let $(\Omega,F, P\times\mu)$ be a probability space and $T>0$ be a fixed time horizon.  \\
\begin{definition}[\textbf{Portfolio Definitions}] 
\begin{itemize}
\item []
    \item Let $R_1, R_2 : \Omega \rightarrow R$ be the return of two stochastic options
    \item Let $ w_1, w_2 = \bm{w} \in [0,1]$ be weights such that $w_1 + w_2 = 1$
    \item Let $R(\bm{w}) = w_1 R_1 + w_2 R_2 - c_{fixed}$ be the return of the portfolio where $c_{fixed}$ is a fixed cost
    \item Risk: $\sigma_P(\bm{w}) = w_1 ^2 Var_P(R_1) + w_2 ^2 Var_P(R_2) + 2 w_1 w_2 Cov_P(R_1,R_2)$
    \item Ambiguity: $\sigma_{\mu}(\bm{w}) = w_1 ^2 Var_{\mu}(R_1) + w_2 ^2 Var_{\mu}(R_2)$
    \item Arrow-Pratt Adjusted Utility: $U(\bm{w}) = -1\times E[R(\bm{w})] - \frac{\lambda}{2} \sigma^2_P (\bm{w}) - \frac{\theta}{2}\sigma^2_\mu (\bm{w})$
\end{itemize}
\end{definition}

\begin{definition}[\textbf{ML Definitions}] 
\begin{itemize}
\item []
    \item Let $MSE_{rec}, MSE_{ref}, MSE_{ret}: \Omega \rightarrow R$ be mean squared errors 
    \item Let $w_{rec}, w_{ref} \in [0,1]$ be weights such that $w_{rec} + w_{ref}  = 1$
    \item Combined MSE: $MSE(\bm{w}) = w_{ref} MSE_{ref} + w_{rec} MSE_{rec} - MSE_{ret}$ 
    \item Sampling Variance: $\sigma^2 _{ML}(\bm{w}) = w_{ref}^2 Var(MSEref) + w_{rec}^2 Var(MSE_{rec}) \\+ 2 w_{ref} w_{rec} Cov(MSEref,MSE_{rec}) $
    \item Model Uncertainty: $\mu^2 _{ML}(\bm{w}) = w_{rec}^2 Var(\Delta_{rec}(t))$
    \item Utility: $U_{ML}(\bm{w}) = -1 \times E[MSE(\bm{w})] - \frac{\lambda}{2} \sigma^2 _{ML} (\bm{w}) - \theta^2 \mu^2 _{ML}(\bm{w})$
\end{itemize}
\end{definition}
We further require the assumptions: \\ 
\textbf{(A1)} Normality: The sampling distributions of $\beta_{rec}$ and $\beta_{ref}$ at time $t$ are normally distributed with $\beta_{rec} \sim N(\mu + \Delta(t), \sigma_{rec})$ , $\beta_{ref} \sim N(\mu, \sigma_{ref})$, where $\Delta(t)$ represents the amount of concept drift.\\
    \textbf{(A2)} Monotonicity: The recalibration error $\Delta(t)$ is increasing monotonically in during times $t \in [0,T]$ \\
    \textbf{(A3)} Fixed cost structure: For fixed budget $c>0$, $w_{ref} \leq \frac{c - c_{model}}{c_{unlab}}$ and $w_{ref} \leq \frac{c}{c_{lab}}$

\begin{theorem} \textbf{Equivalence of Utiliy }
    Under assumptions \textbf{(A1) - (A3)}, there exists a bijective mapping $\bm{\phi}$ between portfolio and ML decision spaces such that:
    \begin{enumerate}
        \item $\bm{\phi}(R_1) = MSE_{ref}$
        \item $\bm{\phi}(R_2) = MSE_{rec}$
        \item $\bm{\phi}(\sigma^2(R_1)) = Var{MSE_{ref}}$
        \item $\bm{\phi}(\sigma^2(R_2)) = Var{MSE_{rec}}$
        \item $\bm{\phi}(\mu(R_1)) = 0$
        \item $\bm{\phi}(\mu(R_2)) = Var(\Delta_{rec})$
    \end{enumerate}
    Under this mapping, $w^* = \argmax U_P(w)$ if and only if $\phi(w^*) = \argmax U_{ML}(\phi(w))$
\end{theorem}

\begin{proof}
First, we show the equivalence of the utility functions. Given any $\bm{w} \in [0,1]$ such that $w_1 + w_2 = 1$:
    \begin{align*}
        U(\bm{w}) &= -1 \times (w_1 E[R_1] + w_2 E[R_2] - c_{fixed})\\
        & - \lambda^2/2 (w_1^2 Var_P(R_1) + w_2 ^2 Var_P(R_2) + w_1 w_2 Cov(R_1, R_2)) \\
        & - \theta^2/2 (w_1^2 Var_{\mu}(R_1) + w_2 ^2 Var_{\mu}(R_2))
    \end{align*} Similarly, under mapping $\bm{\phi}$
    \begin{align*}
        U_{ML}(\bm{w}) &= -1 \times (w_{ref} E[MSE_{ref}] + w_{rec} E[MSE_{rec}] - MSE_{ret})\\
        & - \lambda^2/2 (w_{ref}^2 Var_P(MSE_{ref}) + w_{rec} ^2 Var_P(MSE_{rec}) + w_{ref} w_{rec} Cov(MSE_{ref}, MSE_{rec})) \\
        & - \theta^2/2 ( w_{rec} ^2 Var(\Delta(t)))
    \end{align*}
    Note how the structural form is preserved with the corresponding terms. For both weights: $w_1 + w_2 = 1, w_1, w_2 \geq 0 $ and $w_{rec} + w_{ref} =1, w_{rec} , w_{ref} \geq 0 $. Assumption \textbf{(A3)} also directly maps onto the portfolio position limits. Finally, by \textbf{(A1)}, the optimization of $U_{ML}(\bm{w})$ and $U(\bm{w})$ have unique solutions due to convexity. Thus by first-order conditions: $\frac{d U}{d w_1} = 0 \Leftrightarrow \frac{d U_{ML}}{d w_{ref}}$ $\frac{d U}{d w_2} = 0 \Leftrightarrow \frac{d U_{ML}}{d w_{ret}}$
\end{proof}